\documentclass[draft,showpacs,showkeys,eqsecnum,nofootinbib,aps]{revtex4}

\def\beq{\begin{equation}}
\def\eeq{\end{equation}}
\def\bea{\begin{eqnarray}}
\def\eea{\end{eqnarray}}
\def\nn{\nonumber}

\def\pa{\partial}

\begin{document}
\title{BRST extension of the Faddeev model}
\author{Soon-Tae Hong}
\email{soonhong@ewha.ac.kr}
\affiliation{Department of Science
Education, Ewha Womans University, Seoul 120-750, Korea}
\affiliation{Department of Physics, University of Tokyo, Tokyo 113-0033, Japan}
\author{Antti J. Niemi}
\email{niemi@teorfys.uu.se}
\affiliation{Department of Theoretical Physics,
Uppsala University, P.O. Box 803, S-75108 Uppsala, Sweden}
\affiliation{Department of Physics, University of Tokyo, Tokyo 113-0033, Japan}
\date{\today}%
\begin{abstract}
The Faddeev model is a second class constrained system. Here we construct
its nilpotent BRST operator and derive the ensuing manifestly BRST invariant
Lagrangian. Our construction employs the structure of St\"uckelberg fields
in a nontrivial fashion.
\end{abstract}
\pacs{11.10.Ef; 11.10.Lm; 11.30.-j} \keywords{Hamiltonian
approach; nonlinear model; BRST symmetry; ghost field} \maketitle


The Faddeev model is a natural extension of the Heisenberg
O(3)-model.  It appears in many physical applications from high
energy physics~\cite{niemi} to condensed matter physics~\cite{babaev},
and its prominent feature is the presence of knotted solitons~\cite{niemi2}.
The model is defined by the Lagrangian~\cite{faddeev},
\begin{equation}
L_{0}=\int d^{3}x\left[m^{2}(\pa_{\mu}n^{a})(\pa^{\mu}n^{a})
+\frac{1}{e^{2}}H_{\mu\nu}H^{\mu\nu}\right]. \label{lag}
\end{equation}
Here $m$ is a mass scale and $e$ is a dimensionless coupling
constant, and $H_{\mu\nu}$ are defined as
$$
H_{\mu\nu}=\epsilon_{abc}n^{a}\pa_{\mu}n^{b}\pa_{\nu}n^{c},
$$
where the components $n^{a}$ ($a$=1,2,3) define a vector field
with unit length. Since time derivatives appear in (\ref{lag}) at
most quadratically, the Faddeev model allows for a Hamiltonian
interpretation. But due to the condition
$$n^{a}n^{a}-1= 0.
$$
it is a {\it second class} constrained Hamiltonian system. In
order to maintain manifest Lorentz invariance in the Hamiltonian
formalism, we then need to resort to an appropriate extension of
the Hamiltonian Becchi-Rouet-Stora-Tyutin (BRST)
formalism~\cite{brst}.

We start by interpreting the Lagrangian (\ref{lag}) in terms of
its Hamiltonian variables. From (\ref{lag}),
we find that the canonical momenta conjugate to
the real scalar fields $n^{a}$ are given by
\beq
\pi^{a}\ = \ \frac{\delta L_0 }{\delta \partial_0 n^a}
\ = \ 2m^{2}\partial_{0}{n}^{a}-\frac{4}{e^{2}}A_{i}^{a}A_{i}^{b}
\partial_{0}{n}^{b},
\label{momenta} \eeq where the $A_{i}^{a}$ are
$$
A_{i}^{a}=\epsilon_{abc}n^{c}\partial_{i}n^{b}.
$$
From (\ref{momenta}) we can then solve for the time derivative
$\partial_0 n^a$ in terms of the canonical momenta $\pi^{a}$. The
result can be expressed in terms of a power series in  $1/e^{2}$,
and the first two terms are $$
\pa_{0}n^{a}=\frac{2}{m^{2}}\pi^{a}+\frac{1}{m^{4}e^{2}}
A_{i}^{a}A_{i}^{b}\pi^{b}+O\left(\frac{1}{e^{4}}\right).
$$
This leads to the canonical Hamiltonian \beq H=\int
d^{3}x\left[\frac{1}{4m^{2}}\pi^{a}\pi^{a}+m^{2}(
\partial_{i}n^{a})(\partial_{i}n^{a})
-\frac{1}{e^{2}}H_{ij}^{2}+\frac{1}{2m^{4}e^{2}}
A_{i}^{a}A_{i}^{b}\pi^{a}\pi^{b}\right]
+O\left(\frac{1}{e^{4}}\right), \label{canH} \eeq where the
canonical variables are subject to the Poisson bracket
$$
\{n^{a}(x),\pi^{b}(y)\} =\delta^{ab}\delta^{3}(x-y).
$$

By implementing the Dirac algorithm \cite{dirac64} we conclude
that together with the identity $$ A_{i}^{a}n^{a}=0, $$ our
Hamiltonian system is subject to the following second class
constraints \bea
\Omega_{1}&=&n^{a}n^{a}-1\approx 0,\nonumber\\
\Omega_{2}&=&n^{a}\pi^{a}\approx 0. \label{const22} \eea With
$\epsilon^{12}=-\epsilon^{21}=1$ this second class constraint
algebra is
$$\Delta_{kk^{\prime}}(x,y)=\{\Omega_{k}(x),\Omega_{k^{\prime}}(y)\}
=\epsilon^{kk^{\prime}}n^{a}n^{a}\delta^{3}(x-y).
$$

Following the Hamiltonian quantization scheme for constrained
systems~\cite{dirac64,faddeev86,bft,niemi88,hong02pr} we proceed to convert the
second class constraints $\Omega_i=0$ $(i=1,2)$ into first class
ones. For this we introduce two canonically conjugate
St\"uckelberg fields $(\theta, \pi_{\theta})$ with Poisson bracket
$$
\{\theta(x), \pi_{\theta}(y)\}=\delta^{3}(x-y).
$$
The strongly involutive first class constraints
$\tilde{\Omega}_{i}$ are constructed as a power series of the
St\"uckelberg fields, and the result is
\begin{eqnarray}
\tilde{\Omega}_{1}&=&\Omega_{1}+2\theta,  \nonumber \\
\tilde{\Omega}_{2}&=&\Omega_{2}-n^{a}n^{a}\pi_{\theta}.
\label{1stconst}
\end{eqnarray}

We proceed to the construction of first class canonical variables
$\tilde{{\cal F}} =(\tilde{n}^{a},\tilde{\pi}^{a})$, that
correspond to the original variables ${\cal F}=(n^{a},\pi^{a})$ in
the extended phase space. These variables are obtained as a power
series in the St\"uckelberg fields $(\theta,\pi_{\theta})$, by
demanding that they are in strong involution with the first class
constraints (\ref{1stconst}), that is
\[
\{\tilde{\Omega}_{i}, \tilde{{\cal
F}}\}=0.
\]
After some straightforward but tedious algebra, we obtain for the
first class canonical variables
\begin{eqnarray}
\tilde{n}^{a}&=&n^{a}\left(\frac{n^{c}n^{c}
+2\theta}{n^{c}n^{c}}\right)^{1/2},\nonumber\\
\tilde{\pi}^{a}&=&\left(\pi^{a}-n^{a}\pi_{\theta}\right)
\left(\frac{n^{c}n^{c}}{n^{c}n^{c}+2\theta}\right)^{1/2},\nonumber\\
\tilde{H}_{ij}&=&\epsilon_{abc}n^{a}\pa_{i}n^{b}\pa_{j}n^{c}
\left(\frac{n^{d}n^{d}+2\theta}{n^{d}n^{d}}\right)^{3/2},\nonumber\\
\tilde{A}_{i}^{a}&=&\epsilon_{abc}n^{c}\pa_{i}n^{b}
\frac{n^{d}n^{d}+2\theta}{n^{d}n^{d}}.
\nonumber
\end{eqnarray} Note in particular that now these first class
variables are not truncated but exact, unlike in the case
of the explicit Hamiltonian that we have displayed in
(\ref{canH}).

In terms of the first class variables we obtain for the (exact)
Hamiltonian
$$ \tilde{H}=\int
d^{3}x\left[\frac{1}{4m^{2}}\tilde{\pi}^{a}\tilde{\pi}^{a}
+m^{2}(\partial_{i}\tilde{n}^{a})(\partial_{i}\tilde{n}^{a})
-\frac{1}{e^{2}}\tilde{H}_{ij}^{2}
+\frac{1}{2m^{4}e^{2}}\tilde{A}_{i}^{a}
\tilde{A}_{i}^{b}\tilde{\pi}^{a}\tilde{\pi}^{b}\right].
$$
Explicitly, in terms of the original fields \bea
\tilde{H}&=&\int
d^{3}x\left[\frac{1}{4m^{2}}\left(\pi^{a}-n^{a}\pi_{\theta}\right)
\left(\pi^{a}-n^{a}\pi_{\theta}\right)\frac{n^{c}n^{c}}{n^{c}n^{c}+2\theta}
+m^{2}(\partial_{i}n^{a})(\partial_{i}n^{a})
\frac{n^{c}n^{c}+2\theta}{n^{c}n^{c}}\right.\nonumber\\
& &\left.-\frac{1}{e^{2}}H_{ij}^{2}
\left(\frac{n^{c}n^{c}+2\theta}{n^{c}n^{c}}\right)^{3}
+\frac{1}{2m^{4}e^{2}}A_{i}^{a}A_{i}^{b}\left(\pi^{a}-n^{a}\pi_{\theta}\right)
\left(\pi^{b}-n^{b}\pi_{\theta}
\right)\frac{n^{c}n^{c}+2\theta}{n^{c}n^{c}}\right]. \label{hct}
\eea Notice that this Hamiltonian is strongly involutive with the
first class constraints,
\[
\{\tilde{\Omega}_{i},\tilde{H}\}=0.
\]
Note also that the first class constraints (\ref{1stconst}) can be
rewritten as
\begin{eqnarray}
\tilde{\Omega}_{1}&=&\tilde{n}^{a}\tilde{n}^{a}-1,  \nonumber \\
\tilde{\Omega}_{2}&=&\tilde{n}^{a}\tilde{\pi}^{a}.
\nonumber
\end{eqnarray}
These have the same functional form as
the second class constraints (\ref {const22}) but now we have
the {\it first class} constraint algebra
\[
\{\tilde{\Omega}_{i},\tilde{\Omega}_{j}\}=0.
\]

However, when we now consider the time evolution of the constraint algebra,
as determined by computing the Poisson brackets of the constraints
with the Hamiltonian (\ref{hct}), we conclude from the Poisson bracket
\[
\{ \tilde{\Omega}_{1}, \tilde H \}=0,
\]
that there is a need to improve the Hamiltonian into the following,
equivalent first class Hamiltonian,
$$
\tilde{H}^{\prime}=\tilde{H}+\int d^{3}x~
\frac{1}{2m^{2}}\pi_{\theta}\tilde{\Omega}_{2}.
$$
Indeed, this improved Hamiltonian generates the constraint algebra
\begin{eqnarray}
\{\tilde{\Omega}_{1},\tilde{H}^{\prime}
\}&=&\frac{1}{m^{2}}\tilde{\Omega}_{2},
\nonumber\\
\{\tilde{\Omega}_{2},\tilde{H}^{\prime}\}&=&0.
\nonumber
\end{eqnarray}
Obviously, since the Hamiltonians $\tilde{H}$ and
$\tilde{H}^{\prime}$ only differ by a term which vanishes on the
constraint surface, they lead to an equivalent dynamics on the
constraint surface.


We now proceed to the implementation of the covariant
Batalin-Fradkin-Vilkovisky (BFV) formalism~\cite{bfv}.
We start by the construction of the nilpotent BRST operator.
For this, we introduce two canonical sets of
ghost and anti-ghost fields, together with auxiliary fields
$({\cal C}^{i},\bar{{\cal P}}_{i})$, $({\cal P}^{i}, \bar{{\cal
C}}_{i})$, $(N^{i},B_{i})$ $(i=1,2)$. The BRST operator for our constraint
algebra is then simply
$$
Q = \int {\rm
d}^{3}x~({\cal C}^{i}\tilde{\Omega}_{i}+{\cal P}^{i}B_{i}).
$$
We choose the unitary gauge with
$$\chi^{1}=\Omega_{1},~~~
\chi^{2}=\Omega_{2}$$
by selecting the gauge fixing functional
\[
\Psi = \int {\rm d}^{3}x~(\bar{{\cal C}}_{i}\chi^{i}+\bar{{\cal P}}%
_{i}N^{i}).
\]
Clearly,
\[
Q^{2}=\{Q,Q\}=0,
\]
and explicitly $Q$ is the generator of the following infinitesimal
transformations $$
\begin{array}{ll}
\delta_{Q}n^{a}=-{\cal C}^{2}n^{a},
&~~\delta_{Q}\pi^{a}=2{\cal C}^{1}n^{a}+{\cal C}^{2}(\pi^{a}-2n^{a}
\pi_{\theta}),\\
\delta_{Q}\theta={\cal C}^{2}n^{a}n^{a},
&~~\delta_{Q}\pi_{\theta}=2{\cal C}^{1},\\
\delta_{Q}{\cal C}^{i}=0,
&~~\delta_{Q}\bar{{\cal P}}_{i}=\tilde{\Omega}_{i},\\
\delta_{Q}{\cal P}^{i}=0,
&~~\delta_{Q}\bar{{\cal C}}_{i}=B_{i},\\
\delta_{Q}N^{i}=-{\cal P}^{i}, &~~\delta_{Q}B_{i}=0.
\end{array}
$$ Furthermore, we have \bea
\delta_{Q}\tilde{H}&=&\{Q,\tilde{H}\}=0,\nn\\
\delta_{Q}\{Q,\Psi\}&=&\{Q,\{Q,\Psi\}\}=0, \label{qh} \eea where
the second line follows from the nilpotentcy of the charge $Q$.
The ``gauge fixed" BRST invariant Hamiltonian is now given by
$$
H_{eff}=\tilde{H}-\{Q,\Psi\},
$$
with $\tilde{H}$ defined in (\ref{hct}). It is clearly BRST
invariant.

After some algebra which is associated with the evaluation of the
Legendre transformation of $H_{eff}$, we arrive at the following
manifestly covariant BRST improved (quantum) Lagrangian
\begin{equation}
L_{eff}= L_{0} + L_{WZ} + L_{ghost}
\label{lagfinal}
\end{equation}
where $L_{0}$ is given by (\ref{lag}) and
\begin{eqnarray}
L_{WZ}&=&\int d^{3}x~\left[
\frac{2m^{2}}{n^{c}n^{c}}(\partial_{\mu}n^{a})(\partial^{\mu}n^{a}){\theta}
+\frac{1}{e^{2}}H_{\mu\nu}H^{\mu\nu}\left(3
+\frac{6\theta}{n^{c}n^{c}}+\frac{4\theta^{2}}
{(n^{c}n^{c})^{2}}\right)
-\frac{m^{2}}{(n^{c}n^{c})^{2}}\partial_{\mu}
\theta\partial^{\mu}\theta\right],\nonumber\\
L_{ghost}&=&\int d^{3}x~\left[-m^{2}(n^{a}n^{a})^{2}
(B_{2}+2\bar{{\cal C}}_{2}{\cal C}^{2})^{2}
-\frac{1}{n^{c}n^{c}}\partial_{\mu}\theta\partial^{\mu}B_{2}
+\partial_{\mu}\bar{{\cal C}}_{2}\partial^{\mu}{\cal
C}^{2}\right].
\nonumber
\end{eqnarray}
This is our main result, a manifestly covariant version of the Faddeev
model (\ref{lag}) where the variable $n^a$ is now an
{\it unconstrained} variable. Note that in deriving
(\ref{lagfinal}) we have included all the higher order terms of
$1/e^{2}$, that we truncated in displaying (\ref{canH}).  We also note
that the (BRST gauge fixed) effective Lagrangian (\ref{lagfinal})
is manifestly invariant under the following
(Lagrangian) BRST transformation,
$$
\begin{array}{ll}
\delta_{\epsilon}n^{a}=\epsilon n^{a}{\cal C}^{2},
&~~\delta_{\epsilon}\theta=-\epsilon n^{a}n^{a}{\cal C}^{2},\\
\delta_{\epsilon}\bar{{\cal C}}_{2}=-\epsilon B_{2},
&~~\delta_{\epsilon}{\cal C}^{2}=\delta_{\epsilon}B_{2}=0,\\
\end{array}
$$ where $\epsilon$ is an infinitesimal Grassmann valued parameter.
Finally, we note that the  St\"uckelberg field $\theta$ becomes a nontrivial,
propagating field degree of freedom.


In conclusion, we have derived the BRST improved version of the
Faddeev model. It has the advantage, that the explicit implementation of the
second class constraint which enforces the order parameter $n^a$ to
be normalized into unity, can be avoided. In order to obtain the
BRST version of the Faddeev model, we have first employed the  St\"uckelberg
fields to convert the second class constraint algebra into a first
class algebra. In particular, the  St\"uckelberg fields appear in
a nontrivial manner in the Lagrangian BRST version of the Faddeev model.

\vskip 0.5cm \noindent {\bf Acknowledgments} The authors would
like to thank Tohru Eguchi for warm hospitality at University of
Tokyo, where this work was completed. S.T.H. would like to
acknowledge financial support in part from the Korea Science and
Engineering Foundation Grant R01-2000-00015. The research by A.N.
has been supported by a grant from VR (Vetenskapsr\.adet) and by
the STINT Thunberg Fellowship.

\end{document}